\documentclass[a4paper,11pt,oneside]{article}
\usepackage{amssymb}
 
\usepackage[T1]{fontenc}
\usepackage{graphicx}
\usepackage{amsmath}

 \newcommand{\cl}{\centerline}
\newcommand{\bi}{\begin{itemize}}
\newcommand{\ei}{\end{itemize}}
\newcommand{\ben}{\begin{enumerate}}
\newcommand{\een}{\end{enumerate}}

\newcommand{\bfi}{\begin{figure}[hbtp]}
\newcommand{\efi}{\end{figure}}
 
\newcommand{\dr}{\partial}
\newcommand{\beq}{\begin{equation}}
\newcommand{\eeq}{\end{equation}}
\newcommand{\beqar}{\begin{eqnarray}}
\newcommand{\eeqar}{\end{eqnarray}}
 \newcommand{\tensna}{\overline{\overline{\nabla}}}

\begin{document}

\centerline{\bf Dust trapping in inviscid vortex pairs.}
\centerline{\bf Jean-Régis Angilella}


 \centerline
{Nancy-Universit\'e, LAEGO, Ecole Nationale Supérieure de Géologie,}
 \centerline
{ rue du Doyen Roubault, 54501 Vand\oe uvre-les-Nancy, France}

\vskip.5cm
 \centerline
{\bf Abstract}
{ 
The motion of tiny heavy particles transported in a co-rotating vortex pair, 
with or without particle inertia 
and sedimentation, is investigated. The dynamics of non-inertial sedimenting 
particles is shown to be chaotic, under the combined effect of gravity and of the circular displacement of the vortices. This phenomenon is very sensitive to  particle inertia, if any. By using nearly hamiltonian dynamical system theory for the particle motion equation written in the rotating reference frame, 
one can show that small inertia terms of the particle motion equation strongly modify the
Melnikov function of the homoclinic trajectories and heteroclinic cycles of the unperturbed system, as soon as the particle response time is of the order of the settling time (Froude number of order unity). The critical Froude number above which chaotic motion vanishes and a regular centrifugation takes place is obtained from this Melnikov analysis and compared to numerical simulations. Particles with a finite inertia, and in the absence of gravity, are not necessarily centrifugated away from the vortex system. Indeed, these particles can have various equilibrium positions in the rotating reference frame, like the Lagrange points of celestial mechanics, according to whether their Stokes number is smaller or larger than some critical value. An analytical stability analysis 
reveals that two of
 these points are stable attracting points, so that permanent trapping can occur  for inertial particles injected in an isolated co-rotating vortex pair. 
Particle trapping is observed to persist when viscosity, and therefore vortex coalescence, is taken into account.
Numerical experiments at large but finite Reynolds number show that particles can indeed be trapped temporarilly during vortex roll-up, and are eventually centrifugated away once vortex coalescence occurs.
}
 


 

\section{Introduction - Particle motion equation}

%
%
%

The motion of tiny particles in fluid flows has many unexpected features 
which have been studied for decades in various contexts (chemical engineering,  atmospheric dust, plankton transport in the ocean, planetesimal formation, etc.). Even the simplest particle transport model (that is passive heavy non-interacting Stokes particles) coupled to any simple flow model 
(prescribed laminar flow) lead, in general, to non-integrable differential equations with six degrees of freedom for the particle position and velocity components.
The dynamics is therefore very rich and it is not surprising to observe complex motion emerge in a wide variety of natural flows where dropplets, solid grains or even biological
particles are present. 
The non-integrability of the motion equation of these low-Reynolds number particles is due to the gradients of the base flow, and this flow does not need to be very complex for 
chaotic particle trajectories to emerge. Indeed, many simple fluid flows have been shown
  to transport tiny solid particles in a complex manner, and this motivated numerous theoretical or numerical studies
(Maxey \& Corrsin
\cite{Maxey1986} ; Wang, Maxey, Burton \& Stock \cite{Wang1992} ; Mac Laughlin 
\cite{McLaughlin1988} ;   Fung \cite{Fung1997} ; Tsega, Michaelides \& Eschenazi  \cite{Tsega01} ; Rubin, Jones \& Maxey \cite{Rubin1995} ; Druzhinin \cite{Druzhinin1997} ; Haller \& Sapsis \cite{Haller2008}). Also,
 the understanding of the interaction between particles and elementary vortical structures could even help understand turbulent particle 
transport or  turbulence modification (see for example, in this spirit,
Marcu {\it et al.} \cite{Marcu1995},
Davila \& Hunt \cite{Davila2001},  Ferrante \& Elghobashi \cite{Ferrante2003}).

This paper deals with the motion of low Reynolds number heavy
 particles transported in a co-rotating vortex pair, i.e.  the two-dimensional flow induced by two identical point vortices (figure \ref{DessinHuit}). This
   is one of the simplest unsteady flows with a natural periodicity due to the   mutual influence of the vortices.  Gravity, acting in the $(x,y)$ plane, is also taken into account.
This choice has been motivated by the fact that  various analyses of heavy 
particle motion in horizontal mixing layers, submitted to subharmonic forcing,
have shown that vortex pairing
can significantly influence the structure of the particle cloud (Raju \& Meiburg \cite{Raju1995}, Chein \& Chung \cite{Chein1987},
Kiger \& Lasheras \cite{Kiger1995}). In particular, vortex pairing has been observed to increase particle dispersion and to enhance particle homogenization. 
Even though the flows investigated by these authors are much more complex than the one investigated here, we believe that the inviscid model used here can highlight and quantify some complex features of dust motion during vortex pairing.  

 Firstly, we wish to show that non-inertial heavy particles injected in this flow can have chaotic trajectories under the combined effect of gravity and of the rotation of the vortices (section \ref{noinertie}). This effect tends to increase particle mixing. Also, the effect of weak particle inertia on this phenomenon will be discussed (section \ref{effetinertie}).
Secondly, we will study inertial non-sedimenting particles, and show that some of them can be trapped by two attracting points rotating with the vortices (section \ref{trapping}). The former effect (chaotic particle motion) induces strong particle mixing, whereas the latter (particle accumulation) induces  preferential particle concentration and could lead to aggregate formation.
For the  sake of simplicity, we will  neglect the flow modification due to 
the particles, as well as particle interactions.

In this elementary  flow one can easily show that the distance $2d$ between the vortices remains constant,
and that the vortices rotate around the center point (here the point $x=y=0$) with a constant angular velocity
$\Omega = \Gamma / 4 \pi d^2 $, where $\Gamma$ is the circulation of each vortex.
Even though the flow Reynolds number $\Gamma/\nu $ is large, the particle Reynolds number is assumed to be small, due to the small size of the inclusions.
The simplest motion equation of such low-Reynolds number heavy particles, in the frame rotating with the vortices, is
\beq
m_p \frac{d^2 \vec{X}_p}{dt^2} = m_p \vec{g} + 6 \pi \mu a \Big(\vec{V}_f - \frac{d \vec{X}_p}{dt} \Big) - m_p \vec \gamma_e - m_p \vec \gamma_c
\label{eqmvt0}
\eeq
where $a$, $m_p$ and $\mu$ denote the particle radius, the particle mass and the fluid viscosity respectively. The term $\vec \gamma_e = - \Omega^2 \vec X_p$ is the  acceleration of the rotating frame relative to the laboratory frame.  The last term 
appearing in the motion equation is the inertial Coriolis force, where  
 $\vec \gamma_c = 2 \vec \Omega \times  {d \vec{X}_p} / {dt}$ is the Coriolis acceleration.
Gravity acts in the plane perpendicular to vorticity as indicated in Fig. \ref{DessinHuit}.
In non-dimensional form, using $d$ to normalize length scales and $1/\Omega$
to normalize time scales, we get (without renaming the variables) :
\beq
\tau \frac{d^2 \vec{X}_p}{dt^2} = -V_T \vec e_{y0}(t) +  \vec{V}_f(\vec X_p) - \frac{d \vec{X}_p}{dt}   +
\tau \Big(\vec X_p - 2 \vec e_z \times \frac{d \vec{X}_p}{dt} \Big)
\label{eqmvt1}
\eeq
where $\vec X_p(t)$ denotes the particle position at time $t$,
$\tau =  \Omega m_p/(6 \pi \mu a)$ is the non-dimensional response-time of the particle (Stokes number), and $\vec e_z$ is the unit vector along the $z$ axis.
The vector $ \vec e_{y0}(t) = \sin t \, \vec e_x + \cos t \, \vec e_y$ is the upward vertical unit vector of the non-rotating frame. 
$V_T$ is the non-dimensional terminal velocity and is assumed to be small throughout this paper (weakly sedimenting particles) : 
$$
V_T =  \tau  g / \Omega^2 d \ll 1
$$
Also, it will be convenient to introduce the Froude number $ \mbox{Fr}=d \Omega^2/g=\tau/V_T$. 
Because particles are much heavier than the fluid, Eq. (\ref{eqmvt1}) is valid even 
if the Stokes number $\tau$ is not small, as added mass force, pressure gradient of the undisturbed flow, lift and Basset force are negligible.
 
\vskip.2cm\noindent
The fluid velocity field $\vec{V}_f$, in the rotating frame, is steady and reads :
$$
\vec V_f(x,y) = \Big(\frac{\dr \psi}{\dr y}, -  \frac{\dr \psi}{\dr x} \Big),
$$
where the streamfunction is given by $\psi(x,y) = -2\ln| z^2 - 1 | + (x^2+y^2) / 2$ (and $z=x + i\,y$) for the inviscid vortex pair considered here. 
The former term in $\psi$ is the flow induced by the two vortices, and the latter corresponds to the  opposite of the velocity of the rotating frame relative to the laboratory frame. The curl of $\vec V_f$ is equal to $-2 \vec e_z$ everywhere, except at the vortex centres where it is infinite.
The corresponding streamlines  are plotted in Fig. \ref{DeuxTraj}(a) : they take the   classical form of an eight shape around the location of the vortices $(\pm 1,0)$.
This phase portrait has three hyperbolic saddle points located at $(0,0)$ and
A =
$(\sqrt 5,0)$ and B = $(-\sqrt 5,0)$. The   point $(0,0)$ has two homoclinic trajectories, whereas the two other points are related by a set of heteroclinic trajectories forming two heteroclinic cycles. Because this flow is two-dimensional and steady, no chaotic fluid point trajectories are expected to emerge. However, it will be shown in the next sections that the motion of particles  has complex and unexpected features in such a flow.
 
Three situations are examined.
In  section \ref{noinertie} we examine the case  $\tau =0$ and $0<V_T \ll 1$, where inertia is negligible and a chaotic motion takes place under the effect of the unsteady term due to gravity in Eq. (\ref{eqmvt1}). Then we focus on the case  $0 < \tau \sim V_T \ll 1$ and examine the effect of weak particle inertia on chaotic sedimentation (section \ref{effetinertie}). 
The case 
$V_T = 0$ and $\tau > 0$ is treated in section  \ref{trapping} : it will be shown that non-sedimenting inertial particles have equilibrium positions in the rotating frame, two of which are stable. A numerical experiment, taking into account the viscous diffusion of the vortices,  is shown in the last paragraph of section \ref{trapping}.
 
\section{Chaotic  sedimentation of inertia-free particles}
\label{noinertie}

The case of non-inertial sedimenting particles can be addressed by assuming $\tau \to 0$ in the particle motion equation (\ref{eqmvt1}), so that all inertia forces vanish to leading order. In addition, we  assume that $V_T$ is small but non-zero. We then recover the simplest motion equation for heavy particles in fluid~:
\beq
 \frac{d \vec{X}_p}{dt} =  \vec{V}_f(\vec X_p)  - V_T \vec e_{y0}(t)
\label{eqmvt2}
\eeq
which has been widely used so far (see for example Stommel \cite{Stommel1949}).  The particle dynamics  therefore corresponds to a steady two-dimensional flow 
($\vec V_f$), plus a
time-periodic perturbation due to gravity. 
The dynamical system (\ref{eqmvt2}), in the absence of perturbation ($V_T=0$),   has a homoclinic
trajectory and a heteroclinic cycle~: it is therefore tempting to check whether or not a finite perturbation ($0 < V_T \ll 1$) induces a homoclinic bifurcation, leading to chaotic aerosol motion in the vortex system. This point can be readily addressed by calculating the Melnikov function of any of these  separatrices (say, $\Sigma_i$)~: 
$$
M_i(t_0) = -\int_{-\infty}^{+\infty} \frac{d \vec{X}_0}{dt} \times   \vec e_{y0}(t+t_0)   \, dt
$$
where $t_0$ is the starting time of the Poincar\'e section of the perturbed system (i.e. $T$-stroboscopy of the system with $T = 2\pi / \Omega$) and $\vec X_0(t)$ is the solution of the undisturbed system corresponding to a point moving along the separatrix $\Sigma_i$ (see for example Guckenheimer \& Holmes \cite{GH01}), that is~:
\beq
\frac{d \vec{X}_0}{dt} = \vec V_f(\vec X_0(t)) 
\label{X0}
\eeq
If the Melnikov function has simple zeros when $t_0$ varies, then the Poincar\'e application of the system (or one of its iterates) corresponds to 
a horse-shoe map, and the particle dynamics is chaotic in the vicinity of $\Sigma_i$  (Smale-Birkhoff theorem). 
For the homoclinic trajectory $\Sigma_0$ the coordinates $x_0(t)$ and $y_0(t)$ of $\vec X_0$ are
even and odd functions of time respectively (choosing $\vec X_0(0)$ equal to the intersection between $\Sigma_0$ and $Ox$).
 We obtain, after removing the integral of odd functions~:
$$
M_0(t_0) = 
\alpha_0 \, \sin (t_0)  
$$
where 
$$
\alpha_0 =  \int_{-\infty}^{+\infty} \Big(  \dot x_0(t)\, \sin t + \dot y_0(t)\, \cos t  \Big) dt
$$
and $\dot x_0$ and $\dot y_0$ are the coordinates of $d\vec X_0/dt$. 
For the heteroclinic trajectories $\Sigma_1$ and $\Sigma_2$ the coordinates $x_0(t)$ and $y_0(t)$ of $\vec X_0$ are
odd and even functions of time respectively (choosing $\vec X_0(0)$ equal to the intersection between the separatrix and $Oy$.)
 We obtain, once again removing the integral of odd functions~:
$$
M_i(t_0) = 
\alpha_i \, \cos (t_0)  
$$
where 
$$
\alpha_i =  \int_{-\infty}^{+\infty} \Big( - \dot x_0(t)\, \cos t + \dot y_0(t)\, \sin t  \Big) dt.
$$
for $i=1$ and $i=2$. 
In order to obtain the coordinates of $\vec X_0(t)$ we need to solve (\ref{X0}),
 with the appropriate initial conditions specified above,
on $\Sigma_i$.
The equation of the separatrix $\Sigma_i$ reads 
$
-2\ln| z^2 - 1 | + r^2 / 2 = \psi(\Sigma_i),
$
where   $r = |z|$. By injecting this last equation into the undisturbed system (\ref{X0}) we are led to a purely numerical differential system which can be solved numerically to obtain $x_0(t)$ and $y_0(t)$. From these the above integrals  can be calculated and we get $\alpha_0 \approx -1.6$, $\alpha_1 \approx 1.7$ and $\alpha_2 \approx -5.8$.
Hence, the Melnikov functions of the three separatrices have simple zeros. 
We therefore conclude that, as soon as gravity is taken into account, a bifurcation occurs in the particle dynamics and chaotic trajectories exist in the vicinity of the homoclinic trajectory $\Sigma_0$ and of the heteroclinic cycle $\Sigma_1 \cup \Sigma_2$. Numerical solutions of equation (\ref{eqmvt2}) confirm this result~: Fig. \ref{DeuxTraj} shows the trajectory of two particles injected respectively near the separatrix $\Sigma_0$ and near the right-hand-side vortex. When $V_T = 0$ these particles follow the streamlines, since the flow is steady. When $V_T$ increases the particle injected near the separatrix has a very complex trajectory~: it rotates in a chaotic manner around both vortices alternatively, as a consequence of the breaking of  $\Sigma_0$.
In contrast, the other particle, injected in a non-chaotic zone, has a regular trajectory around the same vortex. For higher $V_T$'s the former particle wanders in several places of the domain.
The Poincar\'e sections of 20 particle trajectories, computed over 100 periods, are shown in figure 
\ref{secpoinc}, for $V_T=0.03$. A stochastic zone is clearly visible in the vicinity of  $\Sigma_0$ and $\Sigma_1 \cup \Sigma_2$, in agreement with the Melnikov theory presented in this paragraph.

\section{Effect of inertia on chaotic sedimentation}
\label{effetinertie}

When particle inertia, though small, is no longer negligible compared to gravity effects, that is if~:
$$
0 < \tau \sim V_T \ll 1
$$
i.e. the Froude number $ \mbox{Fr}=d \Omega^2/g$   is of order unity, then  the particle motion equation (\ref{eqmvt1}) can be solved approximately  by looking for an asymptotic solution of the form (Michael \cite{Michael1968}, Maxey \cite{Maxey1987}) 
$
{d \vec{X}_p}/{dt} = \vec{V}_f(\vec X_p) + \tau \vec V^1 + O(\tau^2) ,
$
and keeping $\mbox{Fr} $ as a fixed parameter  as $\tau \to 0$.
We are led to~:
\beq
\frac{d \vec{X}_p}{dt} = \vec{V}_f(\vec X_p) +  \tau \Big[  - \frac{1}{\mbox{Fr}} \, \vec e_{y0}(t) - \tensna \vec V_f . \vec V_f + \vec X_p - 2 \vec e_z \times \vec V_f     \Big] + O(\tau^2)
\label{asympt}
\eeq
Like in the previous section we  recover a non-autonomous dynamical system with  two degrees of freedom (the phase space being the physical space), in the form of a perturbed hamiltonian system. 
The perturbation is now dissipative and contains the gravity term, the inertia term, the centrifugal and the Coriolis forces. 
The Melnikov function of a separatrix $\Sigma_i$ now reads~:
$$
M_i(t_0) = -\frac{1}{\mbox{Fr}} \int_{-\infty}^{+\infty} \frac{d \vec{X}_0}{dt} \times   \vec e_{y0}(t+t_0)   \, dt + m_i
$$
where :
\beq
m_i = \int_{-\infty}^{+\infty} \frac{d \vec{X}_0}{dt} \times \Big( - \frac{d^2 \vec{X}_0}{dt^2}+ \vec X_0(t) - 2 \vec e_z \times \frac{d \vec{X}_0}{dt}  \Big)
\label{mi}
\eeq
and $\vec X_0(t)$, running on $\Sigma_i$, has exactly the same expression as in the previous section (solution of equation (\ref{X0})). Hence, the calculation of the $t_0$-dependent part
of the Melnikov function is straightforward. For the separatrix $\Sigma_0$ the Melnikov function reads :
$$
M_0(t_0)  = \frac{1}{\mbox{Fr}}\,\alpha_0\, \sin(t_0) + m_0
$$
where $\alpha_0 < 0$ has been calculated in the previous section (for $\Sigma_1$ and $\Sigma_2$ simply replace $\alpha_0$ by $\alpha_i$ ($i=1,2$), $m_0$ by $m_i$ 
($i=1,2$) and $\sin$ by $\cos$).
The parameter $m_i$ is a constant, and manifests the effect of particle inertia on the dynamics. The Melnikov function $M_i(t_0) $ has simple zeros if
\beq
\mbox{Fr} < \frac{|\alpha_i|}{|m_i|} = \mbox{Fr}_i,
\label{Frc}
\eeq
leading to chaotic particle dynamics in the vicinity of $ \Sigma_i$.
By injecting the numerical 
 $\vec X_0(t)$  calculated above into Eq. (\ref{mi}), we get, for each separatrix~:
   $m_0 \approx -42$,     $m_1 \approx -26$
and  $m_2 \approx 8$.
Because ${\mbox{Fr}}$, $\alpha_0$ and $\alpha_1$  are of order unity, we conclude that inertia has a huge effect on $\Sigma_0$ and $\Sigma_1$, as it tends to prevent the occurrence of simple
zeros in their Melnikov function. Instead of the chaotic trajectories observed in the inertia-free limit, a regular centrifugation   will take place
in the vicinity of $\Sigma_0$ and $\Sigma_1$. The effect of inertia is less significant for  $\Sigma_2$. These calculations show that as soon as 
$$
\mbox{Fr} >  \max_i \mbox{Fr}_i = \mbox{Fr}_2 \approx 0.7,
$$
the stochastic layer around  $\Sigma_2$ will vanish also.

In order to check this  result we have solved numerically Eq. (\ref{asympt}), for 1000 particles released slightly above  $\Sigma_2$ (near point C of Fig. \ref{DeuxTraj}(a)), with $\tau = 0.005$ and  $\mbox{Fr} \in [0.1,1.5]$ (Fig. \ref{NpsFrVarie}). If a stochastic layer exists around  $\Sigma_2$, then some particles are likely to penetrate inside the zone bounded by $\Sigma_2 \cup \Sigma_1$. These particles are those who are
located  in the lobes formed between every two intersection
point of the unstable manifold $W^u(B)$ (coming out of the hyperbolic point located near point B of Fig. \ref{DeuxTraj}(a)) and the stable manifold  $W^s(A)$  (converging to the hyperbolic point located near A).  We observe on Fig. \ref{NpsFrVarie} that some particles   indeed penetrate inside the zone bounded by $\Sigma_2 \cup \Sigma_1$ for $\mbox{Fr} < \mbox{Fr}_2$.  The detailed shape of the curve for $\mbox{Fr} < \mbox{Fr}_2$
depends on the shape and position of the initial cloud.
As soon as $\mbox{Fr} > \mbox{Fr}_2$  however, the curve is flat and no particle penetrate into  the zone bounded by $\Sigma_1 \cup \Sigma_2$, in agreement with  the Melnikov analysis.  Indeed, when $\mbox{Fr} > \mbox{Fr}_2$ the   Melnikov function   $M_2(t_0)$  is constant and non-zero, so that $W^s(A)$ and $W^u(B)$ 
never intersect, and particles located outside cannot penetrate into the zone bounded by $\Sigma_2 \cup \Sigma_1$.

\section{Trapping of non-sedimenting inertial particles}
\label{trapping}

 When the particle response time is of order unity and gravity effects are small
($V_T \ll \tau = O(1)$),
particles are expected to be centrifugated away from the vortices. Here we show that this is not always the case, and that some particles, in spite of their finite inertia, can be trapped and remain in the vicinity of the vortex system. 
For the sake of simplicity we assume that the gravity term   in the particle motion equation  (\ref{eqmvt1}) can be neglected, so that the fluid velocity field in the rotating frame is now steady.  
By using this simplification we notice that  (\ref{eqmvt1}) has equilibrium points where the centrifugal pseudo-force balances the Stokes drag :
\beq
\vec 0 = \vec V_f(\vec X_p) + \tau \, \vec X_p.
\label{eqequil}
\eeq
If one of these points were asymptotically stable, which is the case if all
 the corresponding eigenvalues have strictly negative real parts,   some particles could be trapped there and remain fixed in the rotating frame. This would imply permanent trapping by the vortex pair. In order to clarify this point we   analyze, 
in the following lines,  the  equilibrium points and their stability.

The equilibrium equation has a trivial solution, $\vec X_p^{(1)} = (0,0)$, for all $\tau$. 
By re-writing equation (\ref{eqequil}) in polar coordinates $(r,\theta)$, and after a few algebra (see appendix \ref{app1}), we
obtain 
four more equilibrium points if $0<\tau< 2-\sqrt 3$ or $\tau > 2+\sqrt 3$.
These points are symmetric with respect to $(0,0)$ and are denoted
$\pm\vec X_{eq}^{(2)}$ and $\pm\vec X_{eq}^{(3)}$.   They correspond to 
$$
\sin 2\theta = \frac{4 \tau}{\tau^2+1}
$$
that is, $\theta_2 = \arcsin (4\tau/(\tau^2+1))\, /2$ (mod $\pi$) and $\theta_3 = \pi/2 - \theta_2$ (mod $\pi$). The coordinate $r$ of these points is given by $r^2 = (\sin 2 \theta) / \tau + \cos 2 \theta$, that is :
\beq
r^2 = \frac{1}{\tau^2+1}(4 \pm \sqrt{\tau^4-14 \tau^2 + 1})
\label{req}
\eeq
The stability of   these points
 can be addressed by re-writing the motion equation
 (\ref{eqmvt1}), with $V_T=0$, as a standard dynamical system with four degrees of freedom, with
state variable $\vec Y = (x,y,\dot x,\dot y)$, where $(x,y)$ denote the coordinates of the particle. Then  (\ref{eqmvt1}) is an autonomous system
of the form $d\vec Y/dt = \vec F(\vec Y)$. 
One can check, by studying   the roots of the characteristic polynomial 
of the Jacobian matrix $\nabla\mathbf{F}=(\dr F_i/\dr Y_j)$ at the equilibrium points, that   
$\pm\vec X_{eq}^{(2)}$ is always
 unstable  (see appendix \ref{AppStab1}). 
\vskip.2cm
\noindent
Let us now consider the equilibrium point $ \vec X_{eq}^{(3)}$. The determinant 
of  $\nabla\mathbf{F}$ at  $\vec X_{eq}^{(3)} $ is :
\beq
\Delta(\vec X_{eq}^{(3)} ) = -\frac{1}{4 \tau^2}\left(  \tau^4 - 14 \tau^2 + 1 - 4 ( \tau^4 - 14 \tau^2 + 1 )^{1/2}       \right)
\label{Delta4}
\eeq
It is negative only if $\tau^4 - 14 \tau^2 + 1 > 16$, that is $\tau > \sqrt {15}$.
Hence, by applying the same arguments as before, we conclude that for  $\tau > \sqrt {15}$ this equilibrium point is also unstable. However,
for smaller $\tau$, and in particular 
for all $\tau$ in the range $0<\tau< 2-\sqrt 3$, we have    $ \Delta(\vec X_{eq}^{(3)} ) > 0$ : the equilibrium point $ \vec X_{eq}^{(3)}$ is not necessarily unstable.  
To prove that this point is indeed stable we need to solve for the characteristic polynomial of $\nabla\mathbf{F}$ at $ \vec X_{eq}^{(3)}$, which reads~:
$$
P(\lambda) = \lambda^4 + \frac{2 \lambda^3}{\tau} + (2+\frac{1}{\tau^2}) \lambda^2 +  \frac{2 \lambda}{\tau}
 + \Delta
$$
where $\Delta(\tau)$ simply denotes $\Delta(\vec X_{eq}^{(3)})$. By setting $z = \lambda + 1/(2 \tau)$ we are led to a simpler polynomial~:
$$
P(\lambda) = \tilde P(z^2) = \frac{16\tau^4 z^4 + 8\tau^2 (4\tau^2-1) z^2  + 8\tau^2 (2\tau^2\Delta(\tau)-1) + 1 }{16 \tau^4}
$$
which can be readily solved. We obtain~:
$$
z^2_{\pm} = \frac{1-4\tau^2 \pm 4\tau^2 \sqrt{1-\Delta(\tau)}}{4 \tau^2}
$$
Because we consider the case $\Delta > 0$,   two cases emerge : $0< \Delta < 1$ and $\Delta > 1$. 
In the former case ($0< \Delta < 1$) then $z^2_{\pm}$ are real. 
Clearly,
we have $z^2_{-} < 1/ (4 \tau^2)$. If $z^2_- < 0$ 
then  $z_{-}$ is pure imaginary and the two corresponding $\lambda$'s have a negative real part equal to $-1/2\tau$. If  $z^2_- > 0$  then $z_- < 1/(2 \tau)$, so $\lambda < 0$. The second pair of roots are given by :
$$
z^2_{+} = \frac{1 + 4\tau^2 (\sqrt{1-\Delta(\tau)} -1)}{4 \tau^2} < 1/(4 \tau^2)
$$
since $\sqrt{1-\Delta(\tau)} -1 < 0$. Here also we conclude that the corresponding eigenvalues of $\nabla\mathbf{F}$ are real negative (or are complex with negative real parts if $z^2_{+} < 0$).

\vskip.1cm
\noindent
We now turn to  the case $\Delta > 1$. By setting $a=Re(z)$ we are led to
$ a^2 = (Re(z^2)+|z^2|)/2$ and :
\beq
4\tau^2 a^2 = (1-4 \tau^2 + (1-8\tau^2+16 \tau^4 \Delta(\tau))^{1/2})/2
\label{a2}
\eeq
One can prove that the function of $\tau$ appearing at the rhs of this equation is always smaller than 1 (see appendix \ref{app2}). Therefore, $a < 1/(2\tau)$, and the
corresponding eigenvalues of  $\nabla\mathbf{F}$ satisfy $Re(\lambda) = a - 1/(2\tau) < 0$. 
\vskip.2cm
 
We therefore conclude that, for all $\tau$ in the range $0<\tau<2-\sqrt 3$, there exists
a pair of asymptotically stable points located at $\pm \vec X^{(3)}_{eq}$.

\vskip.1cm
 In order to check these results, and to illustrate the effect of  the attracting
points $\pm \vec X_{eq}^{(3)}$ on the particle dynamics, we have computed the trajectories
of 10000 particles injected at $t=0$ in the square $[-3,3]^2$, for $\tau=0.2$. Results are shown in figure \ref{NuagesTauNonNulVT0} where the particle cloud is plotted. 
Graph  (a)  shows that trapping indeed occurs, and that some particles converge
to the points  $ \pm\vec X^{(3)}_{eq}$ predicted by the theory. Also,
graph (b) shows the initial position of all particles such that $| \vec X_p \pm \vec X^{(3)}_{eq} | < 0.2$ at $t=20$ : it gives an approximate view of a 2D cut, in the $(x,y$) plane, of the basin of attraction of the equilibrium points $ \vec X^{(3)}_{eq}$ and $ -\vec X^{(3)}_{eq}$.

In addition, we have conducted a series of runs, where 2000 
particles were injected at $t=0$ all over the
square $[-3,3]^2$, for various response times $\tau$. We then have computed the percentage of particles lying inside the disk of radius 2 and centre (0,0) at long times (here $t=50$).
Most of these particles are those which have been "trapped" by the two attracting points, if any. Clearly, in the absence of attracting points $p(\tau) \approx 0$,
whereas $p(\tau)$ is finite if attracting points are present. So, if $\tau$ is not too small this percentage is a good indicator of the occurrence of trapping. Figure \ref{Nps}
shows that, as soon as $\tau > 2 - \sqrt 3$, the percentage is identically 0, in agreement
with the stability analysis~: the determinant is negative there, so that the point
$\pm \vec X_{eq}^{(3)}$ is unstable.

\vskip.5cm
\noindent
{\it Comparison with a numerical experiment.}
 These results  are valid for inviscid fluids. When viscosity is finite 
the flow is significantly different since vortices coalesce and finally form a single vortex. Particles will therefore be centrifugated away once coalescence occurs. However, if the diffusive time scale over $d$ is larger than the turnover time $\Omega^{-1}$, one can expect particles to be trapped temporarilly in the vicinity of $\pm\vec X_{eq}^{(3)}$
(the stable points of the inviscid flow investigated above) during the interaction of the vortices, even though the equilibrium points do not exist strictly speaking since the relative flow is no longer steady. 
We have therefore performed a series of runs with a two-dimensional (spectral) Navier-Stokes solver including a lagrangian particle tracking algorithm, within a periodic box $[-\pi d,\pi d]^2$. The initial condition corresponds to a pair of Lamb-Oseen vortices 
$\vec V_f(x,y) = (-\omega(r) y,\omega(r) (x-x_v))$ with
$$
\omega = \Gamma  [1-\exp(-3 r^2/\delta^2)]/(2\pi r^2)
$$
where $r^2 = (x-x_v)^2+y^2$ and $x_v = \pm d$ is the initial $x$-coordinate
 of the vortices. The circulation of a single vortex of this kind, placed in an infinite domain, is independent of the core thickness $\delta$ 
and is equal to $\Gamma$.
The fluid and particle motion equations are set non-dimensional by using $\Omega^{-1} = 4 \pi d^2 /\Gamma$
for times and $d$ for scales, like for the inviscid calculations presented above. 
The initial core-size $\delta$ of the vortices is taken to be one tenth of the flow domain. 
The Reynolds number $Re = \Omega d^2/\nu$ is equal to 800.
The number of Fourier modes is 256 in each direction, and a second-order Adams-Bashforth algorithm is used for time stepping. 
 Particles are passive (they do not modify the flow), and do not interact. They are injected at random  at $t=0$ and cover the whole vortex pair. Their initial
velocity is equal to that of the fluid,
 and their trajectories are calculated by solving equation
(\ref{eqmvt1}) (without the inertia forces since the numerical solver used here considers the dynamical equations in the laboratory reference frame). 

Figure \ref{RotSolRe800Tau0.1BIS} shows the evolution of the vortices and of the particle cloud, when $\tau = 0.1$ (this value has been chosen because it 
corresponds to the peak of figure \ref{Nps}). We observe that particle trapping indeed occurs, and persists until $t\approx 12$ (that is about two turns).  
Because this calculation is done
in a periodic domain, the influence of the neighbouring vortex pair located in boxes $[-\pi + 2 n \pi  ,\pi + 2 n \pi  ]^2,\, n \in \mathbf{Z}$ is significant here. However, these vortices do not prevent the appearance of the attracting points predicted by the inviscid theory in infinite domain (section \ref{trapping}).  When coalescence begins, the set of trapped particles
is elongated into a thin filament  (Fig. \ref{RotSolRe800Tau0.1BIS})
which will then be centrifugated away. The dashed line in figure  \ref{RotSolRe800Tau0.1BIS} is the trajectory of the relative equilibrium points  $\pm \vec X_{eq}^{(3)}$ of  the inviscid theory (equation (\ref{req})).
This line  is close to the position of the accumulation points,
in spite of the effect of viscosity and of the neighbouring vortices.  

\section{Conclusion and open questions}

We have investigated the motion of tiny heavy particles with or without inertia 
and gravity in a co-rotating vortex pair. 
For sedimenting inertia-free particles 
($\tau = 0$ and $0 < V_T \ll 1$) we have shown that, under the combined effect
of gravity and of the rotation of the vortices, a chaotic particle dynamics can take place. By writing the particle motion equation in the rotating 
reference frame attached to the vortices, we observed that 
the particle dynamics take  the form of a hamiltonian system submitted to a periodic forcing due to gravity.
A "stretch-sediment and fold" mechanism is   responsible for this chaotic particle dynamics, like the one described in Ref. \cite{Angilella2008}, as gravity plays a central role in the folding of volume elements of the dispersed phase.

Chaotic motion is very sensitive to particle inertia when   $0 < \tau \sim V_T \ll 1$.
Indeed, by using nearly hamiltonian dynamical system theory for the particle motion equation written in the rotating reference frame, 
we have shown that small
inertia terms of the particle motion equation strongly modify the
Melnikov function of the homoclinic trajectory of the unperturbed system. In particular, the stochastic layer in the vicinity  of the separatrices $\Sigma_i$ ($i$=0,1,2)  vanishes as soon as the Froude number is larger than some critical values  $\mbox{Fr}_i$  given by equation (\ref{Frc}). 
A regular centrifugation therefore takes place as soon as the Froude number is above $\max \mbox{Fr}_i \approx 0.7$. Numerical results   confirm this value (figure \ref{NpsFrVarie}).

Particles with a finite inertia, and in the absence of gravity  ($\tau =O(1)$ and $V_T = 0$), can have various equilibrium positions in the rotating reference frame, according to whether their Stokes number is smaller or larger than some critical values of order unity. 
We have rigorously shown that two of these points are stable attracting points, so that permanent trapping occurs  for inertial particles injected in an isolated co-rotating vortex pair. Numerical computations confirm this result, and show that the basin of attraction of the attracting points covers a non-negligible part of the flow domain.
Our analytical calculations also 
show that trapping should stop if $\tau$ exits the range $0<\tau<2-\sqrt 3$. This
point has been confirmed by computing the percentage of trapped particles (which can be thought of as a measure of the basin of attraction), for various reponse times $\tau$ :
this percentage drops to 0 as soon as $\tau > 2 - \sqrt 3$. Note that the points
$\pm\vec X_{eq}^{(2)}$ and $\pm\vec X_{eq}^{(3)}$ also exist for $\tau > 2+\sqrt 3$. However,
the former has been shown to be unstable for all $\tau$, and the latter is also unstable
as soon as $\tau > \sqrt {15}$.
Numerical experiments taking into account the effect of viscosity have been conducted.
We observe that, when the Reynolds number $\Omega d^2/\nu$ is large, the effects of viscosity are sufficiently slow to enable particle trapping in the vicinity of two points, the position of which is in qualitative agreement with the results of the inviscid theory. Once vortex coalescence is complete, particles are centrifugated away.

Note that this effect could be of interest also in the context of particle motion in protosellar gaseous disks. Indeed, large scale anticyclonic vortices might 
play a key role in dust trapping, with the help of the Coriolis force due to the disk rotation, and these vortices are known 
to interact and coalesce. The dynamics of dust during vortex pairing in protostellar  disks could therefore be a topic of interest. 
 
When $\tau =O(1)$ and $V_T$ is non-zero the flow equation is unsteady in the rotating frame, and the attracting equilibrium points described above no longer exist. The analytical treatment of particle dynamics in this case is much more complex, but numerical simulations, not shown here,
 suggest that trapping could exist anyway. Further analyses should clarify this point.

The generalization of these results to vortex pairs with different strengths is the next step of this work. 
Preliminary numerical simulations show that particle trapping still exists in the asymetric case, but the analytical treatment is heavier and   is currently under study. 
Also, when vortices are located in the vicinity of a wall and move  parallel to it,   particle  
 trapping is expected to persist. Indeed,
Vilela \& Motter \cite{Vilela2007} have  observed
attracting points in their simulation of particle transport
in a leapfrogging vortex system (which   also corresponds to a 2D inviscid vortex pair in the vicinity of a wall). We therefore conclude that the attracting points  persist   in the presence of the wall, even though these points have a more complicated trajectory.  A
 detailed theoretical analysis
 would enable to determine the range of parameters leading to such a trapping.

\newpage
\appendix

\section{Equilibrium positions of inertial particles}
\label{app1}
\noindent
The equilibrium equation (\ref{eqequil}) reads
\beqar
\label{eq1}
\frac{-4 y (r^2+1)}{r^4+1+2 (y^2-x^2)} + y + \tau x &=& 0,\\
\frac{ 4 x (r^2-1)}{r^4+1+2 (y^2-x^2)} - x + \tau y &=& 0
\label{eq2}
\eeqar
where $r^2=x^2+y^2$. By using polar coordinates, and combining these two equations, we are led to :
\beq
r^2 = (\sin 2 \theta) / \tau + \cos 2 \theta,
\label{r}
\eeq
together with the admissibility condition required to avoid a null denominator :
\beq
r^4 + 1 - 2 r^2 \cos 2 \theta \not = 0 
\label{admis}
\eeq
By injecting this $r$ into (\ref{eq1})-(\ref{eq2}) we get :
$$
\sin 2\theta \,(\tau \sin\theta-\cos\theta)\left[ (1+ \tau^2 ) \sin 2\theta -  4 \tau
\right] = 0
$$
and
$$
\sin 2\theta \,(\sin\theta+\tau\cos\theta)\left[ (1+ \tau^2 ) \sin 2\theta -  4 \tau
\right] = 0
$$
Because $\tau \sin\theta-\cos\theta $ and $ \sin\theta+\tau\cos\theta$ cannot be both zero, two
  families of solutions emerge. The first one is $\sin 2\theta = 0$ and $r=1$ : it does not satisfy the admissibility condition (\ref{admis}), and must be rejected.
The second one exists  
  if $1+ \tau^2 > 4\tau$, that is $0 < \tau < 2-\sqrt 3$ or $\tau > 2+\sqrt 3$, 
and corresponds to :
$$
\sin 2\theta = \frac{4\tau}{1+ \tau^2},
$$
that is $\theta_0 = \arcsin (\frac{4\tau}{1+ \tau^2})/2$, and
$\theta = \pi + \theta_0$,
 and  $\theta = \pi/2 -  \theta_0$, and $\theta = 3 \pi/2 -  \theta_0$ (mod $\pi$).
These angles corresponds to 4 equilibrium positions in the physical plane, denoted
$\pm\vec X_{eq}^{(2)}$ and $\pm\vec X_{eq}^{(3)}$.
The $r$ coordinates of these points are obtained by using (\ref{r}).

\section{Instability of equilibrium point (2).}
\label{AppStab1}

The determinant $\Delta(\vec X_{eq}^{(2)})$ of  $\nabla\mathbf{F}$ at  $\vec X_{eq}^{(2)} $ is :
$$
\Delta(\vec X_{eq}^{(2)} ) = -\frac{1}{4 \tau^2}\left(  \tau^4 - 14 \tau^2 + 1 + 4 ( \tau^4 - 14 \tau^2 + 1 )^{1/2}       \right)
$$
and one can check, since $2\pm \sqrt 3$ are also roots of $\tau^4 - 14 \tau^2 + 1$,
that  $\tau^4 - 14 \tau^2 + 1$ is strictly positive for all $\tau$ such that 
$0<\tau< 2-\sqrt 3$ or $\tau > 2+\sqrt 3$. We then conclude that  $\Delta(\vec X_{eq}^{(2)} )$ is always negative in this range of $\tau$. 
This implies that the equilibrium point $\vec X_{eq}^{(2)}$ is unstable. 

Indeed, let $\lambda_i$, ($i=1,...,4$) be the roots of the characteristic polynomial of $\nabla\mathbf{F}$ at $\vec X_{eq}^{(2)}$.
Because the coefficients of the polynomial are real these roots are either real or complex conjugate.
 Also $\lambda_1 \lambda_2 \lambda_3 \lambda_4 = \Delta(\vec X_{eq}^{(2)}) < 0$. 
 Clearly, this implies that the four roots cannot be complex conjugates (i.e. $\lambda_2=\bar \lambda_1$ and $\lambda_4=\bar \lambda_3$) since we would have $\Delta(\vec X_{eq}^{(2)}) = |\lambda_1|^2 \, |\lambda_3|^2 > 0$ there. If two of these roots (say $\lambda_1$ and $\lambda_2$) are complex conjugates and the other two are real, then we must have
$ |\lambda_1|^2\lambda_3 \lambda_4  < 0 $, so $\lambda_3 \lambda_4 < 0$ : hence there exists a  strictly positive eigenvalue : the equilibrium point is unstable. Finally, if all the roots are real, then, their product being strictly negative, one of them at least is positive : $\vec X_{eq}^{(2)}$ is unstable.

\section{Upper bound for $Re(z)^2$}
\label{app2}
To prove that $a^2 < 1/(4\tau^2)$ we need to show that (Eq. (\ref{a2})) :
$$
1-4 \tau^2 + (1-8\tau^2+16 \tau^4 \Delta(\tau))^{1/2} < 2
$$
for all $\tau$ in the range $0 < \tau < 2 - \sqrt 3$.
This is equivalent to :
$$
(1-8\tau^2+16 \tau^4 \Delta(\tau))^{1/2} < 1 + 4 \tau^2
$$
Taking the square of these positive numbers, and after simplifications, we are led to :
$$
\tau^2 \Delta < 1 + \tau^2
$$
From (\ref{Delta4}) we get
 $\Delta = (-q+4 \sqrt q)/(4 \tau^2)$, with $q=\tau^4 - 14 \tau^2 + 1$. The last inequality is therefore equivalent to :
$$
\sqrt q < \frac{q}{4} + 1 + \tau^2
$$
Once again, taking the square of these positive numbers,  we are led to an equivalent expression :
$$
q - (\frac{q}{4} + 1 + \tau^2)^2 < 0
$$
Because $q=\tau^4 - 14 \tau^2 + 1$ we have:
$$
q - (\frac{q}{4} + 1 + \tau^2)^2= [-9-124\tau^2-\tau^4(94 - 20\tau^2)-\tau^8]/16 
$$
and this last quantity, being a sum of negative numbers (since $0 < \tau < 2 - \sqrt 3$), is negative. We therefore conclude that $a^2 < 1/(4\tau^2)$.


 

\listoffigures

\newpage
\bfi
\cl{\includegraphics[height=10cm]{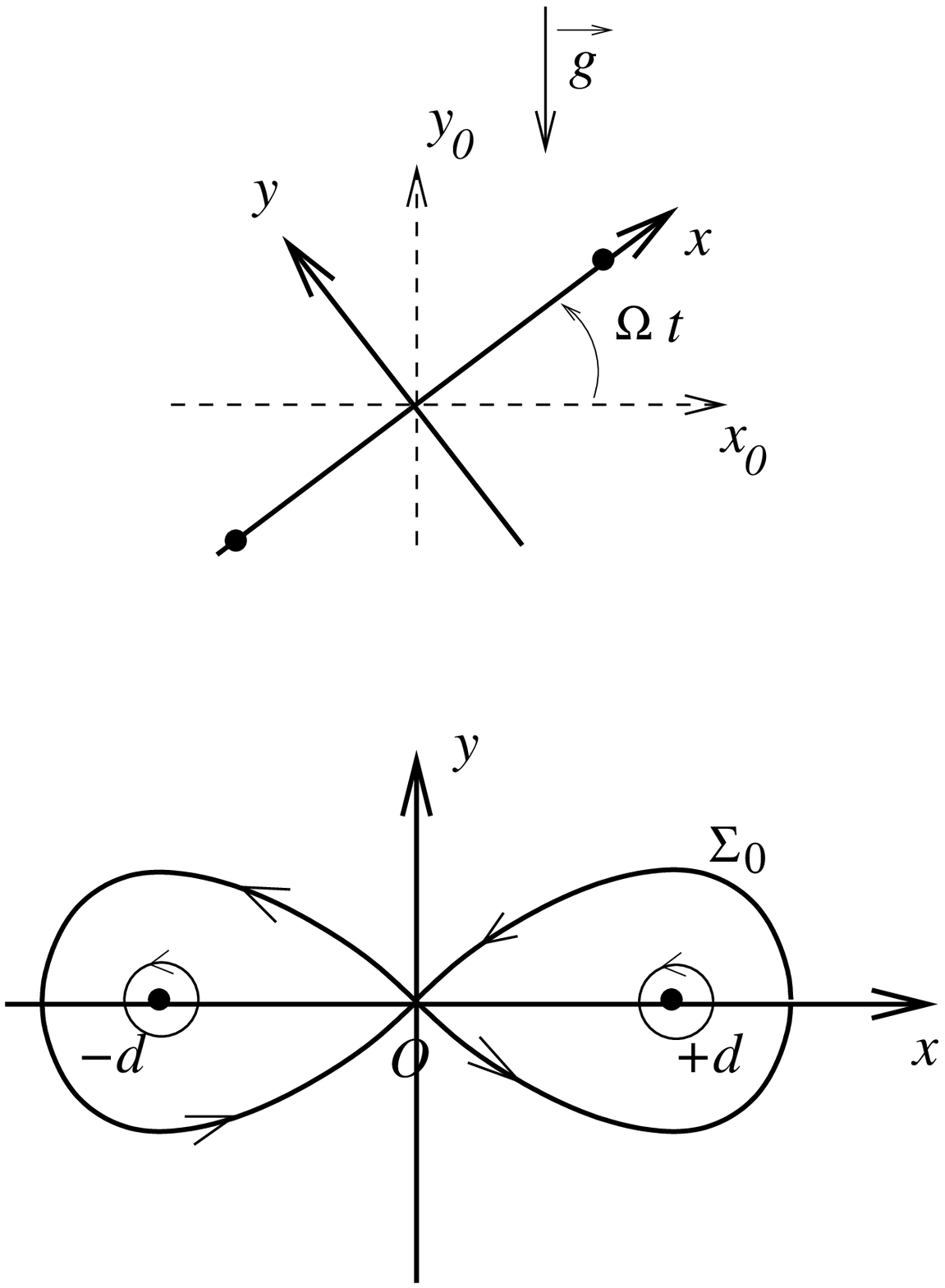}}
\caption{Sketch of the two vortices (black dots), together with the rotating reference frame with axes $(O,x,y)$ and the laboratory frame with axes $(O,x_0,y_0)$. The lower graph is the homoclinic trajectory of the fluid points dynamics in the rotating frame.} 
\label{DessinHuit}
\efi
 \newpage
\bfi
\cl{\includegraphics[height=14cm]{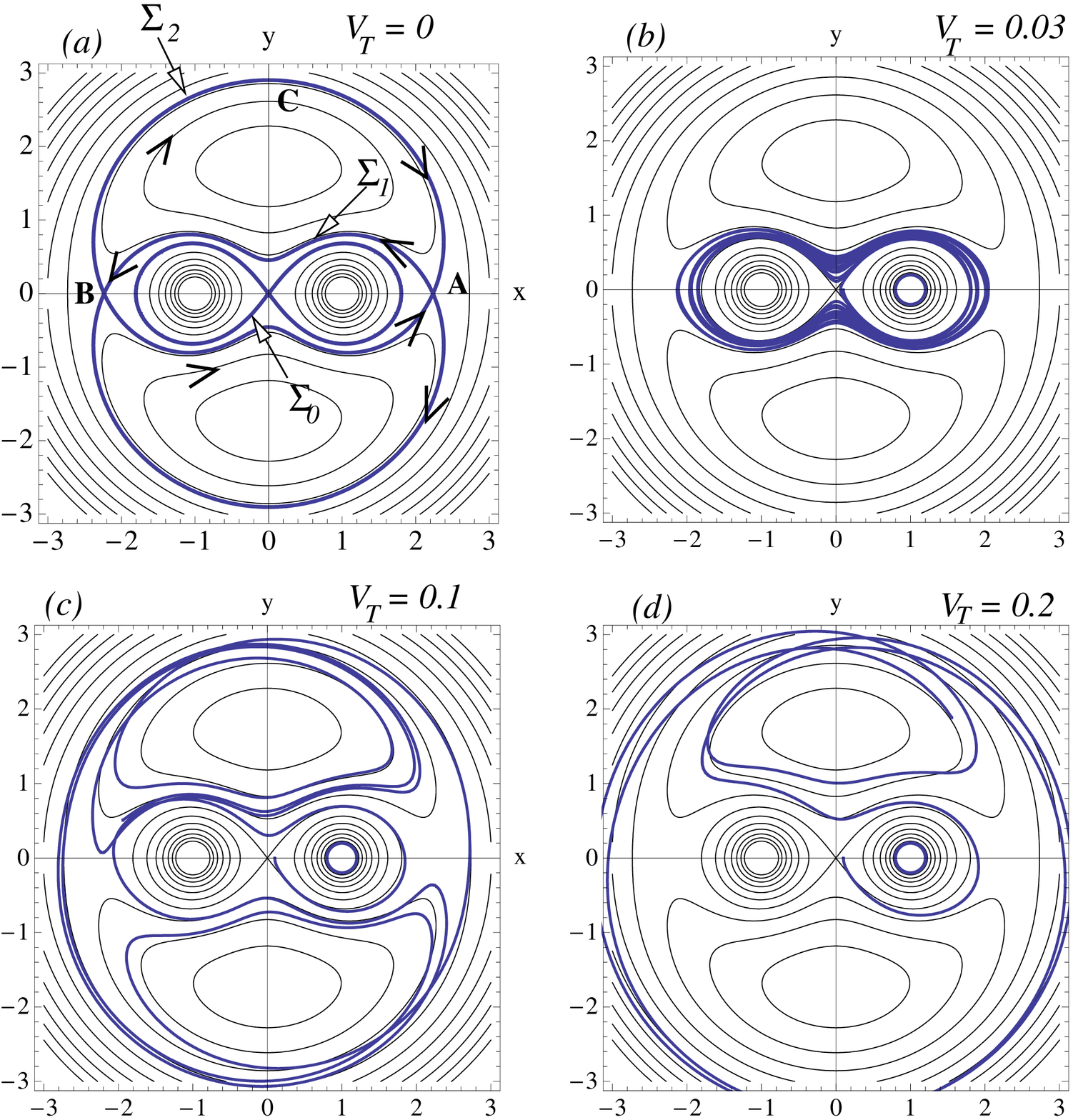}}
\caption{Streamlines in the rotating frame (a). Trajectories of two  inertia-free  particles for various 
terminal velocities (b)(c)(d). The particle injected near the vortex has a regular trajectory around it. In contrast, the particle injected near the separatrix $\Sigma_0$ wanders in various places of the flow domain.  } 
\label{DeuxTraj}
\efi
 
\newpage
\bfi
\cl{\includegraphics[height=10cm]{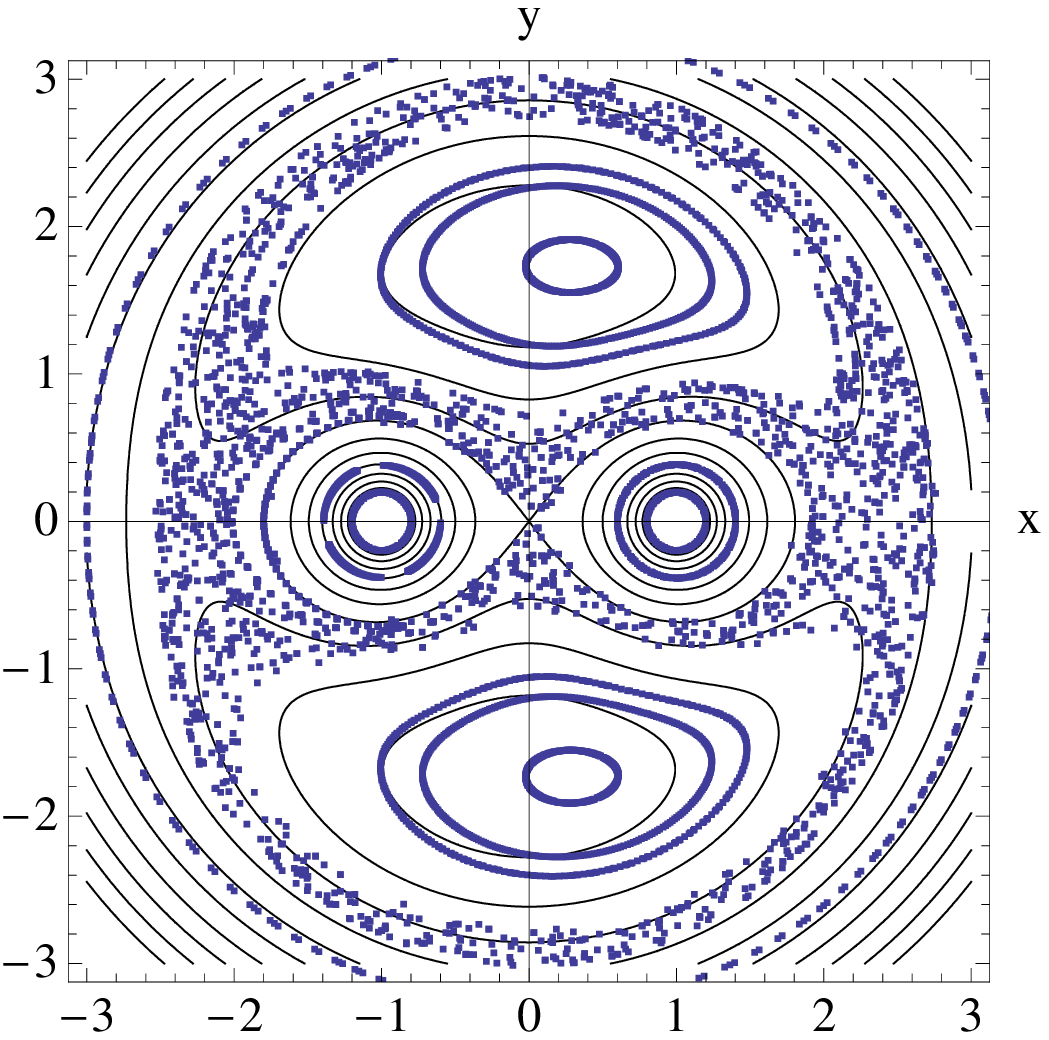}}
\caption{ Poincar\'e section of 20 inertia-free particles, for $V_T=0.03$.} 
\label{secpoinc}
\efi

\newpage
\bfi
\cl{\includegraphics[height=10cm]{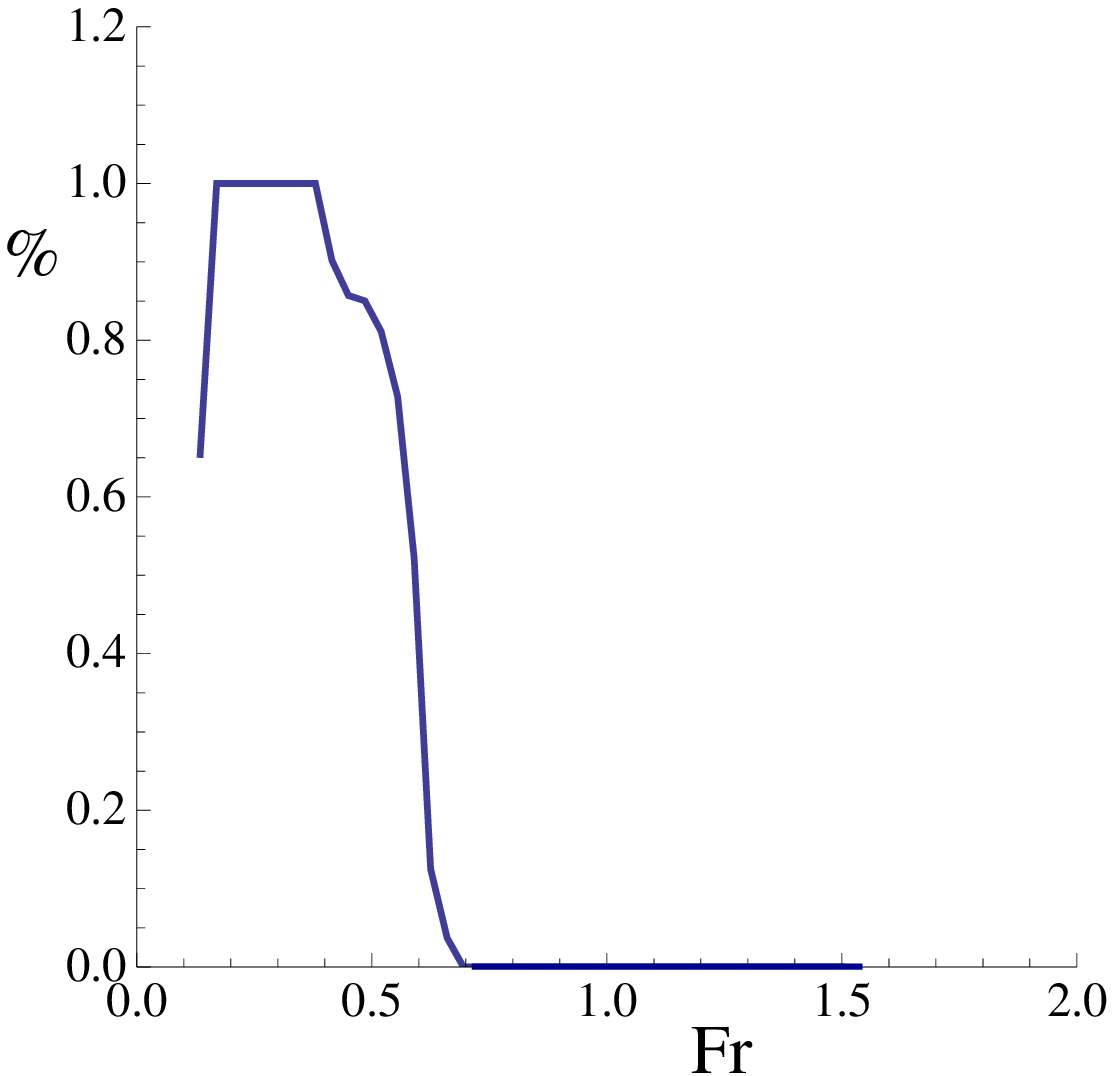}}
\caption{ Plot of the percentage of particles released at $t=0$ slightly above $\Sigma_2$, and crossing $\Sigma_2  $ during the simulation (the final time is $t=100$).
 As soon as $\mbox{Fr} > \mbox{Fr}_2$ the stochastic layer located in the 
vicinity of $\Sigma_2$ vanishes and no particle crosses $\Sigma_2  $, in agreement with  the Melnikov analysis.} 
\label{NpsFrVarie}
\efi

\newpage
\bfi
\cl{\includegraphics[height=7cm]{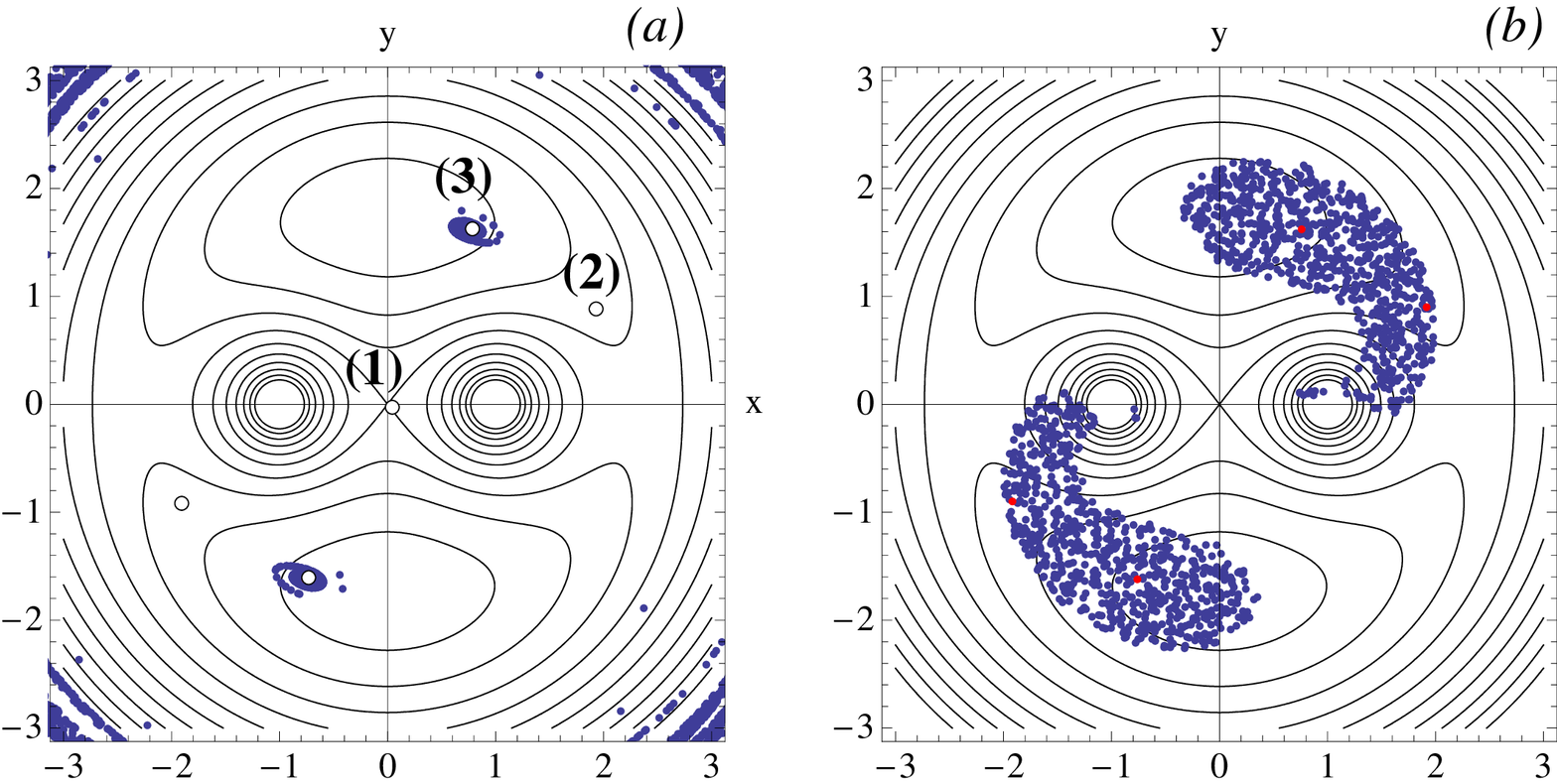}}
\caption{(a) : plot of a particle cloud after 20 convective times. 
At $t=0$ the cloud covers  
 the square $[-3,3]^2$, and is composed of inertial particles with $\tau = 0.2$ and $V_T=0$.  The equilibrium points $ \vec X^{(i)}_{eq}$  
are indicated by the
white dots (i).
Graph (b) shows  the initial position of trapped particles.} 
\label{NuagesTauNonNulVT0}
\efi

\newpage
\bfi
\cl{\includegraphics[height=10cm]{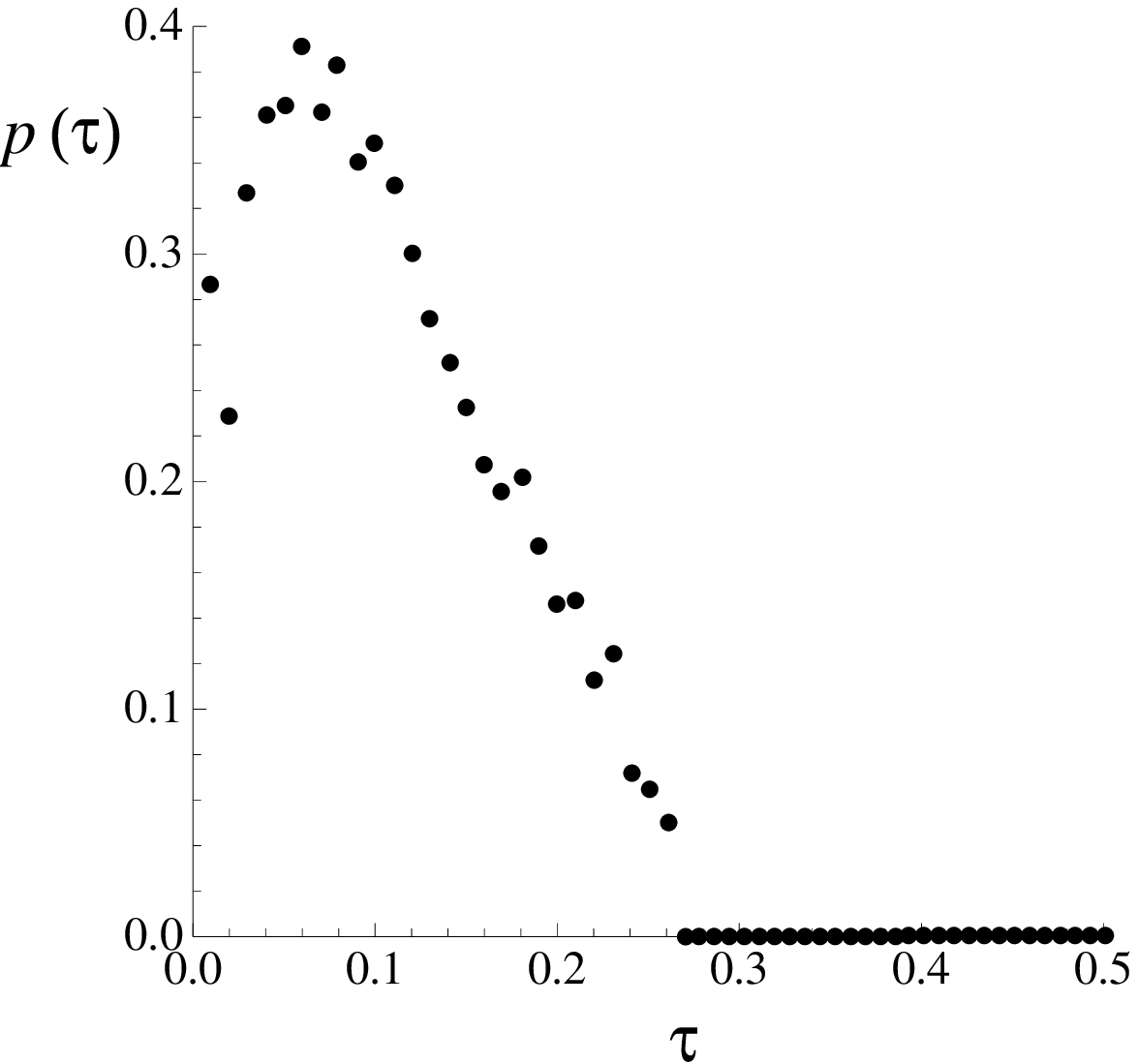}}
\caption{Plot of the percentage of "trapped" particles at $t=50$, versus $\tau$, 
when  $V_T=0$. As soon as $\tau > 2 - \sqrt 3 \approx 0.27$ no trapping is observed ($p(\tau)=0$), in agreement with the stability analysis.} 
\label{Nps}
\efi

\newpage
\bfi
\cl{\includegraphics[height=20cm]{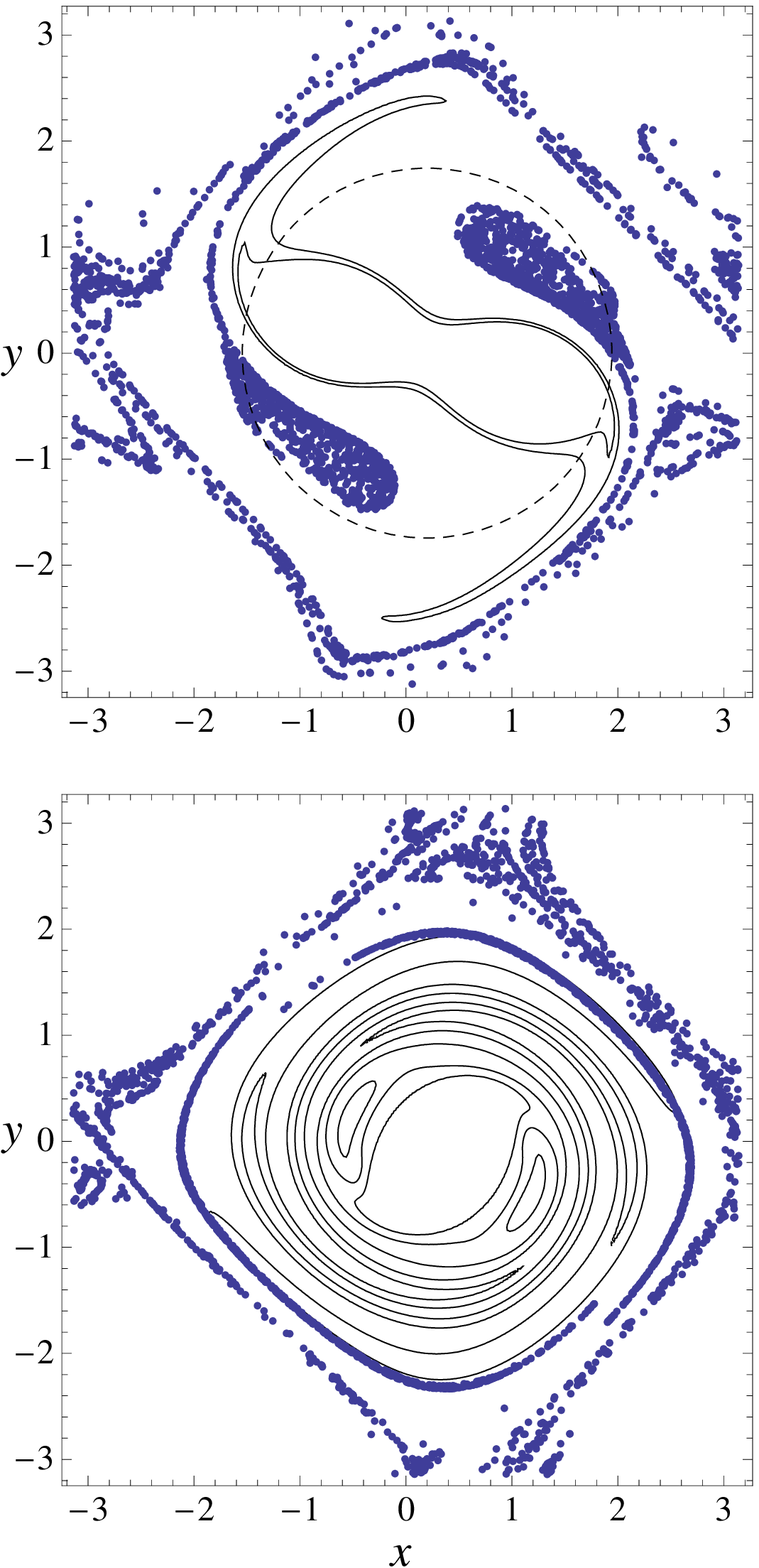}}
\caption{Evolution of a particle cloud  (in the laboratory frame) advected by a vortex pair at large but finite Reynolds number,  for $\tau = 0.1$ and  $V_T=0$. Lines are vorticity contours. Upper figure : $t=10$, lower figure : $t=15$.} 
\label{RotSolRe800Tau0.1BIS}
\efi

\end{document}